\documentclass[referee,showpacs,preprintnumbers,amsmath,amssymb,10pt]{revtex4}

\usepackage{graphicx}
\usepackage{dcolumn}
\usepackage{bm}
\usepackage{hyperref}
\usepackage{color}

\newcommand{\lpl}{\ell_{\mathrm{Pl}}}

\newcommand{\mpl}{M_{\mathrm{Pl}}}

\newcommand{\ptk}{\mathcal{P}_{\mathrm{T}}(k)}

\begin{document}
\title{Inverse volume corrections from loop quantum gravity and the primordial
tensor power spectrum in slow-roll inflation}

\author{J. Grain}
\email{julien.grain@ias.u-psud.fr}
\affiliation{%
Laboratoire AstroParticule \& Cosmologie, Universit\'e Paris 7 Denis Diderot, CNRS, IN2P3\\
10, rue Alice Domon et L\'eonie Duquet, 75205 Paris cedex 13, France\\
{\rm and}\\
Institut d'Astrophysique Spatiale, Universit\'e Paris-Sud 11, CNRS\\ B\^atiments 120-121, 91405 Orsay cedex, France
}%

\author{A. Barrau}%
\email{aurelien.barrau@cern.ch}
\affiliation{%
Laboratoire de Physique Subatomique et de Cosmologie, UJF, INPG, CNRS, IN2P3\\
53, avenue des Martyrs, 38026 Grenoble cedex, France\\
{\rm and}\\
Institut des Hautes Etudes Scientifiques\\
35, route de Chartres, 91440 Bures-sur-Yvette, France
}%

\author{A. Gorecki}%
\email{alexia.gorecki@lpsc.in2p3.fr}
\affiliation{%
Laboratoire de Physique Subatomique et de Cosmologie, UJF, INPG, CNRS, IN2P3\\
53, avenue des Martyrs, 38026 Grenoble cedex, France}

\date{\today}

\begin{abstract}
Together with holonomy corrections, inverse volume terms should be taken
into account when studying the primordial universe in loop quantum cosmology.
We investigate how the tensor power spectrum is modified with respect to the standard
general relativistic prediction by those semiclassical corrections. Depending
on the values of the free parameters of the model, it is shown that the spectrum
can exhibit a very large deviation from its usual shape, in particular with
a very {\it red} slope and a strong running in the infrared limit. 
\end{abstract}


\maketitle

\section*{INTRODUCTION}

A quantum theory of gravity is probably necessary to investigate situations
where General Relativity (GR) breaks down. The early universe is a paradigmatic
example of such a situation where the backward evolution of a classical
space-time inevitably comes to an end after a finite amount of time. Among the
theories willing to reconcile the Einstein gravity with quantum mechanics,
Loop Quantum Gravity (LQG) is especially appealing as it is based on a  
nonperturbative quantization of 3-space geometry (see, {\it e.g.}, \cite{rovelli1} 
and  \cite{smolin1} for an introduction). Loop Quantum Cosmology (LQC) is a 
finite, symmetry reduced model of LQG  suitable for the study of the
whole Universe as a simple physical system (see, {\it e.g.}, \cite{bojo0}).
On the other hand, it is well known that the inflationary scenario is currently the
favored model to describe the first stages of the evolution of the Universe
(see, {\it e.g.}, \cite{linde} for a recent review). Although still debated, 
it has received quite a lot of experimental confirmations, including from the
WMAP 5-Years results \cite{wmap}, and solves most cosmological paradoxes. In
this article, we consider the influence of LQC corrections to general relativity
on the gravitational wave production during inflation. In this intricate
framework, we assume the background to be described by the standard slow-roll
inflationary scenario whereas LQC corrections are taken into account to compute
the propagation of tensor modes. This approach is heuristically justified (to
decouple the physical effects) and intrinsically plausible (as the LQC-driven
superinflation can only be used to set the proper initial conditions to a
standard inflationary stage if the horizon {\it and} flatness problems are both
to be solved \cite{tsuji}). In \cite{grainlqg1,grainlqg2}, holonomy
corrections (due to the fact that loop quantization is based on exponentials of 
the connections, rather than direct connection components) were exhaustively 
considered. We now focus on the other fundamental LQC correction : the inverse
volume --or density-operator-- (due to terms in the Hamiltonian constraint which
cannot be quantized directly but only after being reexpressed as a Poisson bracket
not involving an inverse). In the
first section, the basic formalism is given together with the equations of
motion derived in this framework. The second section deals with the definition of the
power spectra and the  question of initial conditions.
Some analytical results are obtained in the third section. Finally, 
the fourth section explores the full parameter space with numerical
investigations.

\section{Formalism and equations of motion}

Loop Quantum Gravity relies on a Hamiltonian formulation of GR. The
nonperturbative quantum effects associated with the LQG quantization procedure
lead to effective semiclassical LQC equations. The associated hamiltonian
constraint has been obtained by several different approaches
\cite{Bojowald:2004zf,Singh:2005xg,Vandersloot:2005kh,Date:2004zd}. 
The most important step in formulating LQG is to rewrite 
canonical gravity in terms of Ashtekar variables, which are the 
densitized triad $E^a_i$ and the Ashtekar connection $A^i_a$ 
where $E^a_i = e^a_i/|\det e^b_i|$, 
$e^a_i e^b_i = q^{ab}$,
$q_{ab}$ is the spatial metric, 
and $A^i_a = \Gamma^i_a + K^i_a$ 
with 
$\Gamma$ the spin connection and $K$ the extrinsic curvature. The indices run from one to three.
When written in Ashtekar variables, the matter Hamiltonian for a 
general space-time becomes \cite{mulryne}
\begin{equation}
\label{HamPhi}
{\cal H}_\phi = \frac{p_\phi^2 }{2\sqrt{|\det E^c_j|}} +\frac{E^a_iE^b_i\partial_a\phi \partial_b\phi}{2\sqrt{|\det E^c_j|}} + \sqrt{|\det E^c_j|} \, V(\phi)\,~.
\end{equation}

The terms involving inverse expressions 
cannot be straightforwardly quantized and must be regularized by a dedicated
procedure \cite{Thiemann:1996aw,Thiemann:1997rt}.  
The expressions which 
result from this approach are rather complicated, and, in particular, 
are subject to a number of quantization ambiguity parameters.
In principle, the spectrum for the inverse volume can be calculated exactly in
isotropic LQC but the regularization leads to some ambiguities
\cite{Bojowald:2001vw}. The first important scale is set by $\ell_i=\sqrt{\gamma}
\ell_{Pl}$, where  $\gamma$ is the Barbero-Immirzi parameter. It can be understood as 
the  length above which space-time is roughly continuous and the inverse spectrum
can be described by a continuous function. The second relevant scale is 
$\ell_* = \ell_i \sqrt{j/3}$,  where $j$, which takes half integer values, 
is one of the ambiguity parameters.  
Above this scale the eigenvalues of the inverse operator follow the 
classical values, while they radically differ below $\ell_*$.\\

Loop Quantum Gravity introduces strong modifications to the dynamical equations in
the semiclassical regime ({\it i.e.} when the scale factor $a$ is such that
$\ell_{Pl}<a<\ell_*$). They come from the density operator \cite{bojoliv}
\begin{displaymath}
	d_j(a)=\frac{D(q)}{a^3},
\end{displaymath}
with $q=(a/\ell_*)^2$. For the semiclassical universe, the quantum correction factor
is given by \cite{bojopr}
\begin{widetext}
\begin{equation}
D(q)=q^{\frac{3}{2}}\left\{\frac{3}{2m}\left[\frac{1}{m+2}\left(\left|q+1\right|^{m+2}-\left|q-1\right|^{m+2}\right)-\frac{q}{m+1}\left(\left|q+1\right|^{m+1}-\mathrm{sgn}(q-1)\left|q-1\right|^{m+1}\right)\right]\right\}^{\frac{3}{2-2m}},
\end{equation}
\end{widetext}
$m$ being an ambiguity parameter satisfying $0<m<1$.
The cosmological dynamics with a field $\chi$ is then governed by
the following set of differential equations \cite{mielc}~:
\begin{equation}
	\begin{array}{l}
		\displaystyle \ddot{\chi}+\left(3H-\frac{\dot{D}}{D}\right)\dot{\chi}+D\frac{dV}{d\chi}=0 \\
		\displaystyle H^2-\frac{8\pi G}{3}\left[\frac{{\dot\chi}^2}{2D}+V(\chi)\right]=0 \\
		\displaystyle \frac{\ddot{a}}{a}+\frac{8\pi G}{3}\left[\frac{{\dot\chi}^2}{D}\left(1-\frac{\dot{D}}{4HD}\right)-V(\chi)\right]=0,
	\end{array}
\end{equation}
where $H=\dot{a}/{a}$ is the Hubble constant and a {\it dot} means differentiation 
according to the cosmic time. When $a\ge \ell_*$, the Universe enters its classical regime and, in
the limit $a\gg \ell_*$, leading to $D(q)\sim 1$, the usual Klein-Gordon equation is recovered for the 
inflaton field. 
In the semiclassical regime, $a< \ell_*$, it has, however, been shown that
spectacular
modifications to the standard dynamics can be expected. For a scalar field driven
dynamics, it seems that the field can naturally be excited up its self-interaction potential, setting
the initial conditions for slow-roll inflation (see, {\it e.g.},
\cite{tsuji,bojolid,lidsey,mul,tavak,numes,cope}). This is a very appealing feature of
LQC
as the requirement $\phi_i \ge 3 M_{Pl}$ (for simple $\phi^2$ inflationary 
potential), imposed by observations, is rather difficult to set in the standard
framework. Furthermore, a period of superinflation ($\dot{H}>0$) is expected to generically occur 
(see, {\it e.g.}, \cite{bojo3,bojovan,cope2}), irrespectively of the detailed shape of the potential.\\

Following the notation of \cite{cope,cope2}, the equation of motion
for tensor modes with quantum corrections coming from the density operator is given by

\begin{equation}
	\ddot{h}+\left(3H-\frac{\dot{S}}{S}\right)\dot{h}-\frac{S^2}{a^2}\nabla^2h=0,
\end{equation}
We note that such corrections are alternatively denoted by $\alpha$ in \cite{bojo1}.
Because of quantum corrections encoded in the $S$-term, the standard transformation from 
cosmic time $t$ to conformal time $\eta$ ($d\eta=dt/(a(t))$) with the usual field 
redefinition $\Phi=ah$, does {\it not} lead to a Schr\"odinger equation 
(see, {\it e.g.}, Eq. (26) in \cite{mielc}). In particular, there is an 
additional (anti)friction term given by $-\dot{S}/S$. To recast the 
aforementionned differential equation into a Schr\"odinger-like equation, we switch 
from cosmic time to conformal time and redefine the field according to
\begin{displaymath}
	\Phi(\eta,\vec{x})=\frac{a(\eta)}{\sqrt{S(\eta)}}h(\eta,\vec{x}).
\end{displaymath}
Finally, by decomposing the field $\Phi$ over its spatial Fourier modes, one 
obtains
\begin{equation}
	\left[\frac{d^2}{d\eta^2}+S^2(\eta)k^2-V_1(\eta)\right]\phi_k(\eta)=0,
	\label{equmain1}
\end{equation}
with a potential term given by
\begin{equation}
	V_1(\eta)=\frac{a''}{a}-\frac{a'}{a}\frac{S'}{S}+\frac{3}{4}\left(\frac{S'}{S}\right)^2-\frac{1}{2}\frac{S''}{S}.
	\label{potential1}
\end{equation}
The {\it prime} should also be understood as a differentiation according to the conformal time. 
In the classical regime ($S\sim 1$), the potential term becomes $V\approx a''/a$, which is the
usual GR expression.\\

	In addition to density-operator corrections, gravitational waves propagating in a FLRW 
background receive quantum corrections from {\it holonomies} \cite{bojo1}. The
influence of these
LQC corrections has been studied in \cite{grainlqg1,grainlqg2}. 

The values of ambiguity parameter $n$, which depends on the scheme adopted to
quantize holonomies, range between $-1/2$ and $0$, though it was recently shown that $n=-1/2$ is 
favored \cite{corichi}. Moreover, if $n>-1/2$, the holonomy corrections may become a
major contribution to the effective mass of the gravitons at the end of inflation \cite{grainlqg2},
leading to a rather intricate picture.
In the case  $n=-1/2$, the 
potential term reads:
\begin{widetext}
\begin{equation}
	V_2(\eta)=\frac{a''}{a}-2\sqrt{\pi}\gamma\frac{\gamma^2}{\mpl^2}\left[\frac{3}{2\gamma^2L^2(\eta)}\left(1-\sqrt{1-4\frac{H^2(\eta)}{\gamma^2L^2(\eta)}}\right)\right]^2a^{2},
	\label{potential2}
\end{equation}
\end{widetext}
where $L$ is the comoving size of a given patch.

We will resist the temptation to combine $V_1$ and $V_2$ into a single equation which would
describe both holonomy and inverse-volume corrections to the propagation of gravitational waves
as our approach is to decouple, as much as possible, the different physical effects.
This is, however, the next logical step in this study.
Whatever the class of quantum correction considered, the equation of motion for primordial 
gravitons can  be written as a Schr\"odinger-like equation:
\begin{equation}
		\left[\frac{\partial^2}{\partial\eta^2}+E_k(\eta)-V(\eta)\right]f_k(\eta)=0.
	\label{equmain}
\end{equation}
The difference $[E_k(\eta)-V_(\eta)]$ can be seen as the effective squared-frequency 
$\omega^2_k(\eta)$. The correspondence between the different terms is summarized in 
Table~\ref{table1}. If $E_k(\eta)>V(\eta)$, the solution oscillates whereas it becomes a 
coherent sum of an evanescent and an exponentially increasing wave if $E_k(\eta)<V(\eta)$. 
Amplification of quantum fluctuations then arises when $E_k(\eta)<V(\eta)$, {\it i.e.} 
when $\omega_k^2$ is negatively valued. 
\begin{table}
		\begin{tabular}{cc|ccc} \hline\hline
		&& Density-operator && Holonomy \\
		&& corrections && corrections \\ \hline
		&&&& \\
		$f_k$ && $\phi_k\equiv ah/\sqrt{S}$ && $\psi_k\equiv ah$ \\
		$E_k(\eta)$ && $S^2(\eta)k^2$ && $k^2$ \\
		$V(\eta)$ && $V_1(\eta)$ && $V_2(\eta)$ \\ \hline\hline
		\end{tabular}
		\caption{Correspondence between the energylike and potential-like term.}
		\label{table1}
\end{table}
\vspace*{0.3cm}
		
From now on, we switch to our main working hypothesis: as in \cite{grainlqg2}, the background is supposed to be classical
({\it i.e.} described by the usual slow-roll inflationary picture) whereas the mode propagation is
corrected by the inverse-volume LQC term. This makes sense as a phase of standard inflation is
anyway mandatory after the superinflation regime. Furthermore, this allows us to
understand in details the physical origin of the observed features.
During slow-roll inflation, the scale factor is given by
\begin{displaymath}
	a(\eta)=\ell_0\left|\eta\right|^{-1-\epsilon},
\end{displaymath}
with $\epsilon$ the first parameter of the slow-roll expansion. The energy density is assumed to be related to the Hubble parameter via the standard Friedmann 
equation.  The $S$ operator is given  by \cite{bojo1}
\begin{equation}
	S(q>1)\approx1+\lambda q^{-\kappa/2},
	\label{d}
\end{equation}
with $\lambda$ and $\kappa$ two positive constants, not well constrained in
homogeneous models ($\kappa=2n$ with the notation of \cite{bojo1}). 
Setting $\ell_* \sim \ell_{Pl}$ to remain consistent with our hypothesis of a full
"standard" ({\it i.e.} classical)
evolution of the background, one obtains the following energy and potential terms for 
density operator corrections:
\begin{eqnarray}
E_k(\eta)&=&\left[1+2\lambda\left(\frac{\lpl}{\ell_0}\right)^\kappa\left|\eta\right|^{\kappa(1+\epsilon)}\right]k^2,
\label{elqg}\\
V(\eta)&=&\frac{2+3\epsilon}{\eta^2}+\lambda\kappa(1+2\epsilon)\left(\frac{\lpl}{\ell_0}\right)^\kappa\left|\eta\right|^{\kappa(1+\epsilon)-2},
\label{vlqg}
\end{eqnarray}
and, for holonomy corrections: 
\begin{eqnarray}
	E_k(\eta)&=&k^2, \\
V(\eta)&=&\frac{2+3\epsilon}{\eta^2}-2\sqrt{\pi}\gamma^3(1+4\epsilon)\left(\frac{\lpl}{\ell_0}\right)^2\left|\eta\right|^{2\epsilon-2}.
\end{eqnarray}

\section{Analytic results}
The general equation of motion is far too complicated to be analytically solved in the general
case. However, for some particular values of the parameters, the spectrum can be computed, at
least in the infrared (IR) or ultraviolet (UV) limits. Those calculations are both useful by
themselves and convenient to check the numerical results obtained in the next section. 

Throughout all this article, the convention of \cite{martin} is used for the normalization of 
initial states. The particle interpretation of the considered quantum field theory imposes the 
Wronskian $W$ of the mode functions to be equal to $-i$. However, because we are working with 
rescaled quantities, the mode functions have to be normalized so that:
\begin{eqnarray}
	W_\eta(\phi^\dag,\phi)&\equiv&\phi\partial_\eta\phi^\dag-\phi^\dag\partial_\eta\phi \label{wronskian} \\
	&=&-\frac{16i\pi}{\mpl}. \nonumber
\end{eqnarray}
The power spectrum then reads:
\begin{equation}
	\ptk=\frac{2k^3}{\pi^2}\left|\frac{\sqrt{S}\phi_k}{a}\right|^2_{k/aH\to0}.
\end{equation}
As $S\sim1$ at the end of inflation, the power spectrum can be safely approximated by
\begin{equation}
	\ptk=\frac{2k^3}{\pi^2}\left|\frac{\phi_k}{a}\right|^2_{k/aH\to0}.
\end{equation}

	\subsection{$\kappa=1$ and $\epsilon=0$: UV limit}
When $\kappa=1$ and $\epsilon=0$, the UV limit of the power spectrum can be analytically 
computed. The equation of motion can be rewritten as
\begin{equation}
\frac{d^2\phi}{d\eta^2}+\left[\left(1+2Z\left|\eta\right|\right)k^2-\frac{2}{\eta^2}-\frac{Z}{\left|\eta\right|}\right]\phi=0,
\end{equation}
with $Z\equiv\lambda(\lpl/\ell_0)^{\kappa}=\lambda(\lpl/\ell_0)$ in this case.

It is useful to define two regions:  region I is such that the potential 
term can be neglected and region II is such that the time-dependent term in the energy 
can be neglected. Obviously, region I corresponds to $\eta\ll\eta_c$ where $\eta_c$ 
is the potential crossing time defined by $E(\eta_c)=V(\eta_c)$. Region II corresponds to
$-Z^{-1}\ll\eta\leq0$. In region I, the mode functions are given by a linear combination (LC) of Airy functions:
\begin{equation}
	\phi_I(\eta)\sim A_{I}\mathbf{Ai}(x)+B_{I}\mathbf{Bi}(x),
\end{equation}
with
\begin{equation}
	x=-k^{2/3}(2Z)^{-2/3}\left(1+2Z\left|\eta\right|\right).
\end{equation}
The two coefficients are determined by the standard Wronskian condition
\begin{equation}
	\left(2Zk^2\right)^{1/3}\frac{2i\mathrm{Im}\left[A^\dag_IB_{I}\right]}{\pi}=-i\frac{16\pi}{\mpl^2},
\end{equation}
leading to the natural choice
\begin{eqnarray}
	A_I&=&\frac{2\pi\sqrt{2}}{\left(k\sqrt{2Z}\right)^{1/3}\mpl}, \\
	B_I&=&-i\frac{2\pi\sqrt{2}}{\left(k\sqrt{2Z}\right)^{1/3}\mpl}.
\end{eqnarray}
In region II, the mode functions are given by a LC of Coulomb wavefunctions: 

\begin{equation}
	\phi_{II}(\eta)\sim A_{II}F_1(Z/2k,k\left|\eta\right|)+B_{II}G_1(Z/2k,k\left|\eta\right|).
	\label{coulomb-app}
\end{equation}

In the UV regime ({\it i.e.} $k\to\infty$), the potential crossing time is roughly given by 
$\eta_c\propto-1/k$ and region I extends to values of $\eta$ higher than $-1/Z$.
As a consequence, the regions overlap and the matching can be done for conformal times such that 
\begin{displaymath}
	k^{-1}\ll\left|\eta\right|\ll Z ^{-1}.
\end{displaymath}
This region  corresponds to the time when the density-operator is close to unity until 
horizon crossing. In this overlapping region, the Coulomb functions take the following approximate form:
\begin{equation}
	\phi_{II}\sim\left[A_{II}\sin{\left(-k\eta+\varphi_{II}\right)}+B_{II}\cos{\left(-k\eta+\varphi_{II}\right)}\right],
\end{equation}
with 
\begin{displaymath}
	\varphi_{II}=-\frac{\pi}{2}-\frac{\gamma_e
	Z}{2k}+\arctan{\left(\frac{Z}{4k}\right)},
\end{displaymath}
where $\gamma_e$ is the Euler's constant and $\varphi_{II}\sim-\pi/2$ in the UV limit.
On the other hand, still in the UV regime, the Airy functions can be approximated by sine and 
cosine functions:
\begin{eqnarray}
	\lim_{x\to-\infty}\mathbf{Ai}(x)&=&\frac{1}{\sqrt{\pi\sqrt{\left|x\right|}}}\cos{\left(\frac{2}{3}\left|x\right|^{3/2}-\frac{\pi}{4}\right)}, \\
	\lim_{x\to-\infty}\mathbf{Bi}(x)&=&-\frac{1}{\sqrt{\pi\sqrt{\left|x\right|}}}\sin{\left(\frac{2}{3}\left|x\right|^{3/2}-\frac{\pi}{4}\right)}.
\end{eqnarray}
Taking the limit $x\to\infty$ (as $k\to\infty$ and $\eta$ is in the overlapping region) and 
performing a Taylor expansion in $Z\left|\eta\right|$, this leads to
\begin{widetext}
\begin{equation}
	\phi_I\sim\frac{1}{\sqrt{\pi}}\left(\frac{2Z}{k}\right)^{1/6}\left\{\left[A_I\cos{(\varphi_I)}-B_I\sin{(\varphi_I)}\right]\cos{\left(-k\eta\right)}+\left[A_I\sin{(\varphi_I)}+B_I\cos{(\varphi_I)}\right]\sin{\left(-k\eta\right)}\right\},
\end{equation}
with $\varphi_I=k/3Z-\pi/4$.
\end{widetext}

After matching the solutions, one easily obtains the coefficients in region II:
\begin{eqnarray}
	A_{II}&=&\frac{2i\sqrt{2\pi}}{\mpl\sqrt{k}}e^{-i\varphi},\\
	B_{II}&=&iA^\dag_{II},
\end{eqnarray}
where the UV limit of $\varphi_{II}$ has been used. The power spectrum is derived by taking the asymptotic limit for small arguments 
of Eq. (\ref{coulomb-app}). This can be performed by noticing that in the UV 
regime, the Coulomb wave functions are well approximated by Bessel functions. 
This leads to
\begin{equation}
	\ptk=\left(\frac{\lpl}{\ell_0}\right)^2\left(\frac{8\Gamma(3/2)}{\pi}\right)^2,
\end{equation}
which coincides with the power spectrum in GR.

	\subsection{$\kappa(1+\epsilon)=2$: full spectrum and IR \& UV limits}
In this particular case, it is possible to calculate the full power spectrum and to derive
explicit IR and UV limits. Some algebra is, however, necessary.\\

The equation to be solved is the following:
\begin{widetext}
\begin{equation}
\frac{d^2\phi_k}{d\eta^2}+\left\{\left[1+2\lambda\left(\frac{\lpl}{\ell_0}\right)^\kappa\left|\eta\right|^{\kappa(1+\epsilon)}\right]k^2-\frac{2+3\epsilon}{\eta^2}-\lambda\kappa(1+2\epsilon)\left(\frac{\lpl}{\ell_0}\right)^\kappa\left|\eta\right|^{\kappa(1+\epsilon)-2}\right\}\phi_k=0.
\end{equation}
\end{widetext}
With
\begin{eqnarray}
	Z&=&\lambda\left(\frac{\lpl}{\ell_0}\right)^\kappa, \\
	K^2&=&k^2-\lambda\kappa(1+2\epsilon)\left(\frac{\lpl}{\ell_0}\right)^\kappa,
\end{eqnarray}
this can be written as
\begin{equation}
	\frac{d^2\phi_k}{d\eta^2}+\left[K^2+2Zk^2\left|\eta\right|^2-\frac{2+3\epsilon}{\eta^2}\right]\phi_k=0.
\label{ab1}
\end{equation}
To solve this equation, we will express it as a {\it General Confluent Equation} whose general form is
given by
\begin{widetext}
\begin{eqnarray}
	&&\frac{d^2w}{dz^2}+\underbrace{\left[\frac{2A}{z}+2f'+\frac{bh'}{h}-h'-\frac{h''}{h'}\right]}_{\Lambda_1}\frac{dw}{dz}\\
&&~~~~~~~~+\underbrace{\left[\left(\frac{bh'}{h}-h'-\frac{h''}{h'}\right)\left(\frac{A}{z}+f'\right)+\frac{A(A-1)}{z^2}+\frac{2Af'}{z}+f''+{f'}^2-\frac{a{h'}^2}{h}\right]}_{\Lambda_2}w=0, \nonumber
\label{ab2}
\end{eqnarray}
\end{widetext}
and whose solution is
\begin{equation}
	w(z)=z^{-A}e^{-f(z)}\bigg[C_1M\left(a,b,h(z)\right)+C_2U\left(a,b,h(z)\right)\bigg],
\end{equation}
$M,~U$ being Kummer functions and $C_i$ being integration constants.

To write Eq.~(\ref{ab1}) as Eq.~(\ref{ab2}), $\Lambda_1$ must vanish. We define
\begin{eqnarray}
	h(z)&=&\beta z^\alpha \\
	f(z)&=&\frac{h(z)}{2},
\end{eqnarray}
where $\alpha$ and $\beta$ are constants. 
The requirement $\Lambda_1=0$ reads as
\begin{equation}
	2A+\alpha b-\alpha+1=0.
\end{equation}
$\Lambda_2$ must then be equaled to the ($E_k(\eta)-V(\eta)$) term in Eq.~(\ref{ab2}). 
With our choice for $h$ and $f$, $\Lambda_2$ reads:
\begin{equation}
\frac{A(A-1+b\alpha-\alpha+1)}{z^2}-\left(\alpha^2\beta\left(a-\frac{b}{2}\right)-\frac{\alpha^2\beta^2z^{\alpha}}{4}\right)z^{\alpha-2}.
\end{equation}
Taking $\alpha=2$, this can be written as
\begin{equation}
	\Lambda_2=\frac{A(A+2b-2)}{z^2}-4\beta\left(a-\frac{b}{2}\right)-\beta^2z^2,
\end{equation}
which equals $E_k(\eta)-V(\eta)$ if $z=\left|\eta\right|$. Identifying the other terms leads to
\begin{eqnarray}
	z&=&\left|\eta\right|, \\
	\alpha&=&2, \\
	\beta^2&=&-2Zk^2\\
	4\beta\left(a-\frac{b}{2}\right)&=&-K^2, \\
	A(A+2b-2)&=&-2-3\epsilon, \\
	2A+2b-1&=&0.
\end{eqnarray}
This can be easily solved to obtain
\begin{eqnarray}
	A&=&-\frac{1}{2}\pm\frac{3}{2}\sqrt{1+\frac{12}{9}\epsilon} \\
	b&=&1\mp \frac{3}{2}\sqrt{1+\frac{12}{9}\epsilon} \\
	\beta&=&\pm ik\sqrt{2Z}\\
	a&=&\frac{1}{2}\mp\frac{3}{4}\sqrt{1+\frac{12}{9}\epsilon}\mp\frac{K^2}{4ik\sqrt{2Z}}.
\end{eqnarray}

To determine the signs, we require the solution to converge to the usual GR solution when the LQC
terms are vanishing. This means:
\begin{equation}
	\lim_{Z\to0}{\phi_k}=\sqrt{z}\left[D_1J_{\nu}(kz)+D_2Y_{\nu}(kz)\right],
	\label{ab3}
\end{equation}
with
\begin{equation}
	\nu=\frac{3}{2}\sqrt{1+\frac{12}{9}\epsilon}=\mp(b-1).
\end{equation}
When $Z\to0$, one obtains
\begin{equation}
	a\equiv\mp\frac{k}{4i\sqrt{2Z}},
\end{equation}
which tends to $\mp i\infty$. In the limit $|a|\to\infty$, Kummer functions can be rewritten
with Bessel functions:
\begin{eqnarray}
\lim_{Z\to0}\left[M(a,b,h(z))\right]=\Gamma(b)\left(\frac{kz}{2}\right)^{A+1/2}J_{\mp\nu}(kz),
\end{eqnarray}
and $b$ can be reexpressed as a function of either $A$ or $\nu$.
When plugged into the general solution with Kummer functions this leads to
\begin{equation}
	\lim_{Z\to0}{\phi_k}=\sqrt{z}\left[E_1J_{\mp\nu}(kz)+E_2Y_{\mp\nu}(kz)\right]\left[1+\mathcal{O}\left(z^2\right)\right],
\end{equation}
where $E_i$ are constants expressed with $\Gamma$ functions and the $C_i$ coefficients. 
To ensure the equality with Eq.~(\ref{ab3}), the (+) sign must be chosen for $b$ and the ($-$) sign
for A (the sign of $\beta$ is not relevant).
The solution to Eq.~(\ref{ab1}) is therefore
\begin{widetext}
\begin{eqnarray}
	\phi_k(\eta)&=&\frac{\exp{\left(-ik\sqrt{\frac{Z}{2}}\eta^2\right)}}{\left|\eta\right|^{-\frac{1}{2}-\frac{3}{2}\sqrt{1+\frac{12}{9}\epsilon} }}\Bigg[C_1M\left(\frac{1}{2}+\frac{3}{4}\sqrt{1+\frac{12}{9}\epsilon}+\frac{iK^2}{4k\sqrt{2Z}},1+\frac{3}{2}\sqrt{1+\frac{12}{9}\epsilon},ik\sqrt{2Z}\eta^2\right) \\ 
	&&+C_2U\left(\frac{1}{2}+\frac{3}{4}\sqrt{1+\frac{12}{9}\epsilon}+\frac{iK^2}{4k\sqrt{2Z}},1+\frac{3}{2}\sqrt{1+\frac{12}{9}\epsilon},ik\sqrt{2Z}\eta^2\right)\Bigg].
\end{eqnarray}
\end{widetext}
After performing a first-order Taylor expansion in $\epsilon$, the parameters are given by
\begin{eqnarray}
	A&=&-2-\epsilon, \\
	b&=&\frac{5}{2}+\epsilon, \\
	a&=&\frac{5}{4}+\frac{\epsilon}{2}+\frac{iK^2}{4k\sqrt{2Z}}.
\end{eqnarray}

To explicitly derive the power spectrum, the solution is rewritten as
\begin{widetext}
\begin{equation}
		\phi_k(\eta)=\frac{1}{x^A}\exp{\left(-ik\sqrt{\frac{Z}{2}}x^2\right)}\left[C_1M(a,b,ik\sqrt{2Z}x^2)+C_2U(a,b,ik\sqrt{2Z}x^2)\right],
\label{ab9}
\end{equation}
\end{widetext}
where $x=\left|\eta\right|$. The integration constants are determined by the Wronskian condition.
To compute the Wronskian, we will set $h=k\sqrt{2Z}x^2$ and focus on the remote past.
The solution is
\begin{widetext}
\begin{eqnarray}
		\phi_k(h)&=&\left(k\sqrt{2Z}\right)^{A/2}h^{-A/2}e^{-i\frac{h}{2}}\left[C_1M(a,b,ih)+C_2U(a,b,ih)\right], \\
		&\simeq&\left(k\sqrt{2Z}\right)^{A/2}\frac{(i)^{\mathrm{Re}[-a]}}{h^{1/4}}\left\{\lambda\exp{\left(i\frac{h}{2}+i\mathrm{Im}[a]\ln{(h)}\right)}+\mu\exp{\left(-i\frac{h}{2}-i\mathrm{Im}[a]\ln{(h)}\right)}\right\},
\end{eqnarray}
\end{widetext}
with
\begin{eqnarray}
	\lambda&=&e^{-\pi\mathrm{Im}[a]/2}C_1\frac{\Gamma(b)}{\Gamma(a)} \\
	\mu&=&e^{\pi\mathrm{Im}[a]/2}\left[C_2+C_1e^{i\pi a}\frac{\Gamma(b)}{\Gamma(a-b)}\right].
\end{eqnarray}
In the remote pas, the Wronskian can be written as
\begin{equation}
	\sqrt{h}W_h\left\{\phi^\dag_k,\phi_k\right\}\approx-i\left(k\sqrt{2Z}\right)^A\left[\left|\lambda\right|^2-\left|\mu\right|^2\right].
\end{equation}
And the Wronskian condition
\begin{equation}
	\sqrt{h}W_h\left\{\phi^\dag_k,\phi_k\right\}=\frac{8i\pi}{\mpl^2\sqrt{k\sqrt{2Z}}},
\end{equation}
leads to
\begin{equation}
\left|\lambda\right|^2-\left|\mu\right|^2=-\frac{8\pi}{\mpl^2\left(k\sqrt{2Z}\right)^{A+1/2}}.
\end{equation}
Setting $\lambda=0$, this means that
\begin{eqnarray}
	C_1&=&0 \\
	C_2&=&\frac{2\sqrt{2\pi}}{\mpl}\left(k\sqrt{2Z}\right)^{-A/2-1/4}e^{-\pi\mathrm{Im}[a]/2}.
\end{eqnarray}

To obtain the power spectrum, the limit $\eta\to0$ is taken in Eq~(\ref{ab9}):
\begin{eqnarray}
	\phi_k(\eta\to0)&\simeq&C_2\left|\eta\right|^{-A}\frac{\Gamma(b-1)}{\Gamma(a)}\left(ik\sqrt{2Z}\eta^2\right)^{1-b} \\
	&\simeq&C_2\frac{\Gamma(\frac{3}{2}+\epsilon)}{\Gamma(\frac{5}{4}+\frac{\epsilon}{2}+\frac{iK^2}{4k\sqrt{2Z}})}\left(ik\sqrt{2Z}\right)^{-\frac{3}{2}-\epsilon}\left|\eta\right|^{-1-\epsilon} \nonumber \\
&\simeq&C_2\frac{\Gamma(\frac{3}{2}+\epsilon)}{\Gamma(\frac{5}{4}+\frac{\epsilon}{2}+\frac{iK^2}{4k\sqrt{2Z}})}\left(ik\sqrt{2Z}\right)^{-\frac{3}{2}-\epsilon}\frac{a(\eta)}{\ell_0}. \nonumber
\end{eqnarray}
The spectrum is finally given by
\begin{widetext}
\begin{eqnarray}
	\mathcal{P}_{\mathrm{T}}(k)=\left(\frac{\lpl}{\ell_0}\right)^2\frac{16}{\pi}\left|\Gamma\left(3/2+\epsilon\right)\right|^2\left(2Z\right)^{-\epsilon/2-3/4}\frac{k^{3/2-\epsilon}\exp{\left(-\frac{\pi
K^2}{4k\sqrt{2Z}}\right)}}{\left|\Gamma\left(\frac{5}{4}+\frac{\epsilon}{2}+\frac{iK^2}{4k\sqrt{2Z}}\right)\right|^2}.
\end{eqnarray}
\end{widetext}
This establishes the full tensor power spectrum.
In the IR and UV limits, $\mathrm{Im}(a)\to\pm\infty$, so
\begin{equation}	
\left|\Gamma\left(\frac{5}{4}+\frac{\epsilon}{2}+\frac{iK^2}{4k\sqrt{2Z}}\right)\right|^2\simeq2\pi\exp{\left(-\left|\frac{\pi
K^2}{4k\sqrt{2Z}}\right|\right)}\left|\frac{K^2}{4k\sqrt{2Z}}\right|^{3/2+\epsilon}.
\end{equation}
After performing the Taylor expansion in $\epsilon$ and taking the $IR$ and $UV$ values of $K^2$,
one obtains
{\begin{widetext}
\begin{eqnarray}
\mathcal{P}_{\mathrm{T}}^{(IR)}&=&\left(\frac{\lpl}{\ell_0}\right)^2\left(\frac{2^{\frac{3}{2}}\Gamma\left(3/2+\epsilon\right)}{\pi}\right)^2\left[2Z(1+\epsilon)\right]^{-\frac{3}{2}-\epsilon}k^3\exp{\left(\frac{\pi\sqrt{2Z}(1+\epsilon)}{2k}\right)}\label{limir}\\
\mathcal{P}_{\mathrm{T}}^{(UV)}&=&\left(\frac{\lpl}{\ell_0}\right)^2\left(\frac{2^{3+\epsilon}\Gamma\left(3/2+\epsilon\right)}{\pi}\right)^2k^{-2\epsilon}\left[1+(3+5\epsilon)\frac{Z}{k^2}+\mathcal{O}(k^{-4})\right].
\label{limuv}
\end{eqnarray}
\end{widetext}

It can easily be seen from those equations that important modifications to the usual GR picture
are expected in the IR regime whereas the UV behavior is equivalent to the GR 
one up to $\mathcal{O}\left(k^{-2}\right)$. In the
IR range, one can both notice a very red spectrum and a very strong running of the index.\\

The tilt of the spectrum is given by

\begin{eqnarray}
n_{\mathrm{T}}^{(IR)}&=&3-\frac{\pi\sqrt{2Z}(1+\epsilon)}{2k}\label{tiltlimir}\\
n_{\mathrm{T}}^{(UV)}&=&-2\epsilon-(3+5\epsilon)\frac{2Z}{k^{2}},
\label{tiltlimuv}
\end{eqnarray}


and the running by

\begin{eqnarray}
\alpha_{\mathrm{T}}^{(IR)}&=&\frac{\pi\sqrt{2Z}(1+\epsilon)}{2k}\label{runlimir}\\
\alpha_{\mathrm{T}}^{(UV)}&=&(3+5\epsilon)\frac{4Z}{k^{2}}.
\label{tiltlimuv}
\end{eqnarray}


\section{Numerical results}

To explore the full parameter space beyond the particular case $\kappa (1+\epsilon)=2$, there is
probably no other way to go than to perform a full numerical investigation of the problem. To this aim, a
dedicated code was developed. Initial and final conditions, defined respectively in the remote past and at the 
end of inflation, are however required to perform such a computation. 

\subsection{Boundary conditions}
To define the initial states, we first perform the following transformation:
\begin{eqnarray}
	\left|\eta\right|=e^x,&&\phi_k=u(x)e^{x/2}.
\end{eqnarray}
The differential equation (\ref{equmain}) with $E$ and $V$ given by Eq.~(\ref{elqg}) and
Eq.~(\ref{vlqg}) now reads:
\begin{widetext}
\begin{equation}
	\frac{d^2u}{dx^2}+\left[\left(1+2\lambda\left(\frac{\lpl}{\ell_0}\right)^\kappa e^{\kappa(1+\epsilon)x}\right)e^{2x}k^2-\lambda\kappa(1+2\epsilon)\left(\frac{\lpl}{\ell_0}\right)^\kappa e^{\kappa(1+\epsilon)x}-\frac{9}{4}-3\epsilon\right]u=0.
\end{equation}
\end{widetext}
In the remote past, the new variable $x$ tends to infinity and the squared frequency in the 
above Schr\"odinger-like equation becomes dominated by the term proportional to 
$e^{\kappa(1+\epsilon)x+2x}k^2$. With this approximation and using 
$y=[\kappa(1+\epsilon)x+2x]/2$, it becomes a Bessel equation. For $\eta\to-\infty$, the mode functions
can therefore be written as a LC of Hankel functions of order 0:
\begin{equation}
	\phi_k(\eta\to-\infty)=\sqrt{-\eta}\left[A_kH_0(z_\eta)+B_kH^\dag_0(z_\eta)\right],
	 \label{phiini}
\end{equation}
with
\begin{equation}
	z_\eta=\frac{2\sqrt{2\lambda}}{\kappa(1+\epsilon)+2}\left(\frac{\lpl}{\ell_0}\right)^{\frac{\kappa}{2}}k\left|\eta\right|^{[\kappa(1+\epsilon)+2]/2}.
\end{equation}
The amplitudes of the mode functions are determined by requiring the appropriate Wronskian 
condition as defined in Eq. (\ref{wronskian}). Using 
\begin{displaymath}
	W_z(H^\dag_0(z),H_0(z))=\frac{4i}{\pi z},
\end{displaymath}
it can be shown that the Wronskian is given by
\begin{equation}
W_\eta(\phi^\dag,\phi)=-\frac{2i}{\pi}\left[\kappa(1+\epsilon)+2\right]\left[\left|A_k\right|^2-\left|B_k\right|^2\right].
\end{equation}
The natural choice
\begin{eqnarray}
	A_k&=&\frac{2\pi\sqrt{2}}{\left[\kappa(1+\epsilon)+2\right]\mpl}, \\
	B_k&=&0
\end{eqnarray}
will therefore be made to fulfill the Wronskian condition. 

At the very end of inflation, one can notice that whatever the LQC correction considered, 
it becomes subdominant when compared with the GR term. As a consequence,
the equation of motion becomes in this regime
\begin{equation}
	\frac{1}{\phi_k}\frac{d^2\phi_k}{d\eta^2}=\frac{\ddot{a}}{a}.
\end{equation}
The solution is therefore given by a growing and a decaying mode. 
Clearly, to estimate the power spectrum, the growing mode is the most important one:
\begin{equation}
	\phi_k(\eta\to0)\sim C_ka(\eta).
	\label{fonctionfit}
\end{equation}
From this last expression, one simply rewrites the power spectrum as a function of $C_k$ :
\begin{equation}
		\ptk=\frac{2k^3}{\pi^2}\left|C_k\right|^2.
\end{equation}

\subsection{Results}

For the numerical investigations, the solution is first taken as Hankel functions of order 0, given
by Eq.~(\ref{phiini}), in the limit $\eta\to -\infty$. It is then numerically propagated
according to Eq.~(\ref{equmain1}) with a variable step fourth-order Runge-Kutta code. When
$\eta\to 0$, the numerical results are fitted with Eq.~(\ref{fonctionfit}) and the resulting value
of $C_k$ is used to compute the power spectrum. Figure~\ref{fig0} displays, for
$\kappa(1+\epsilon)=2$, both the numerical
results and the analytical calculation in the IR limit obtained with Eq.~(\ref{limir}). The
excellent agreement illustrates the reliability of the approach. Our code has also 
been tested with LQC corrections set to zero: the numerical results are
also in agreement with the analytically known prediction of standard slow-roll 
inflation.

\begin{figure}
	\begin{center}
	\includegraphics[scale=0.45]{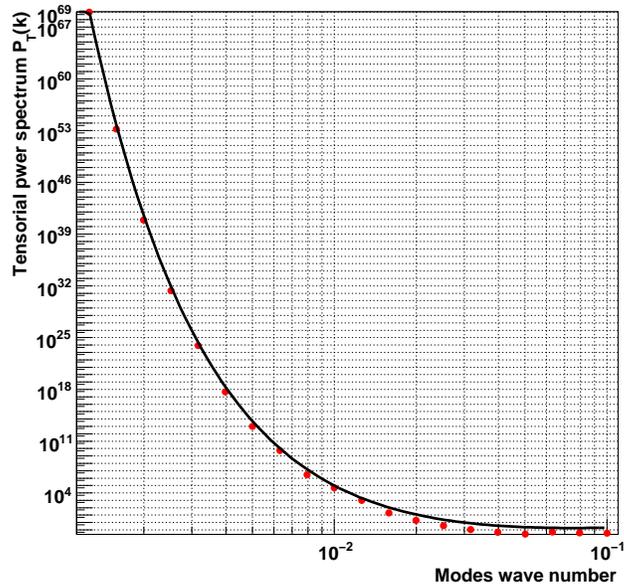}
	\caption{Primordial tensor power spectrum as a function of the wavenumber. The solid line
	is the analytical calculation whereas filled circles are the numerical results.}
	\label{fig0}
	\end{center}
\end{figure}

\begin{figure}
	\begin{center}
	\includegraphics[scale=0.45]{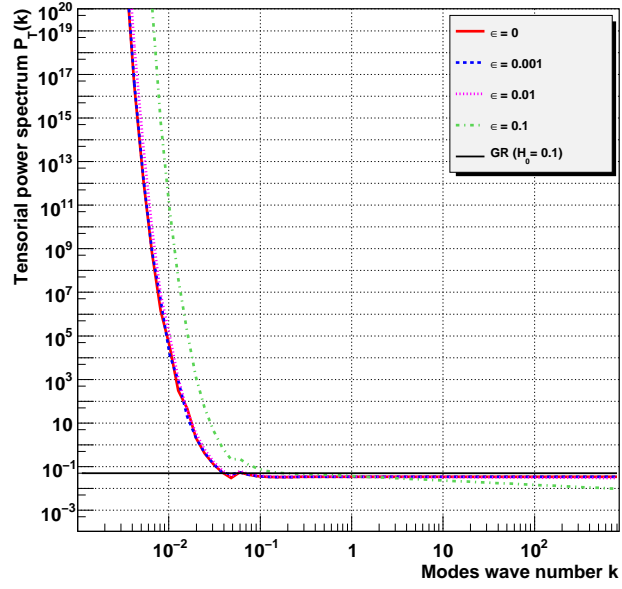}
	\caption{Primordial tensor power spectrum as a function of the wavenumber for different
	values of the slow-roll parameter $\epsilon$.}
	\label{fig1}
	\end{center}
\end{figure}

\begin{figure*}
	\begin{center}
	\includegraphics[scale=0.4]{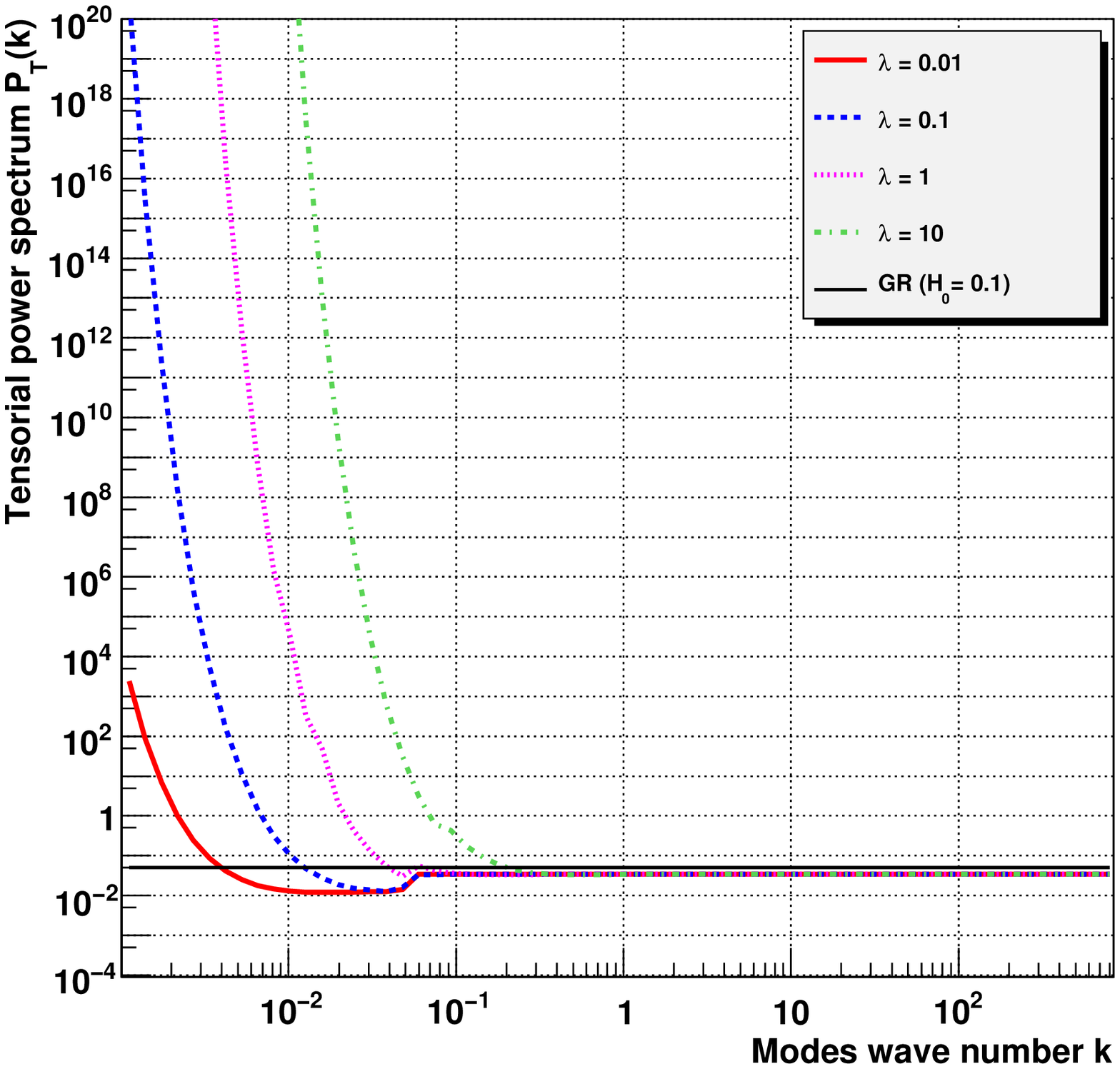}
	\includegraphics[scale=0.4]{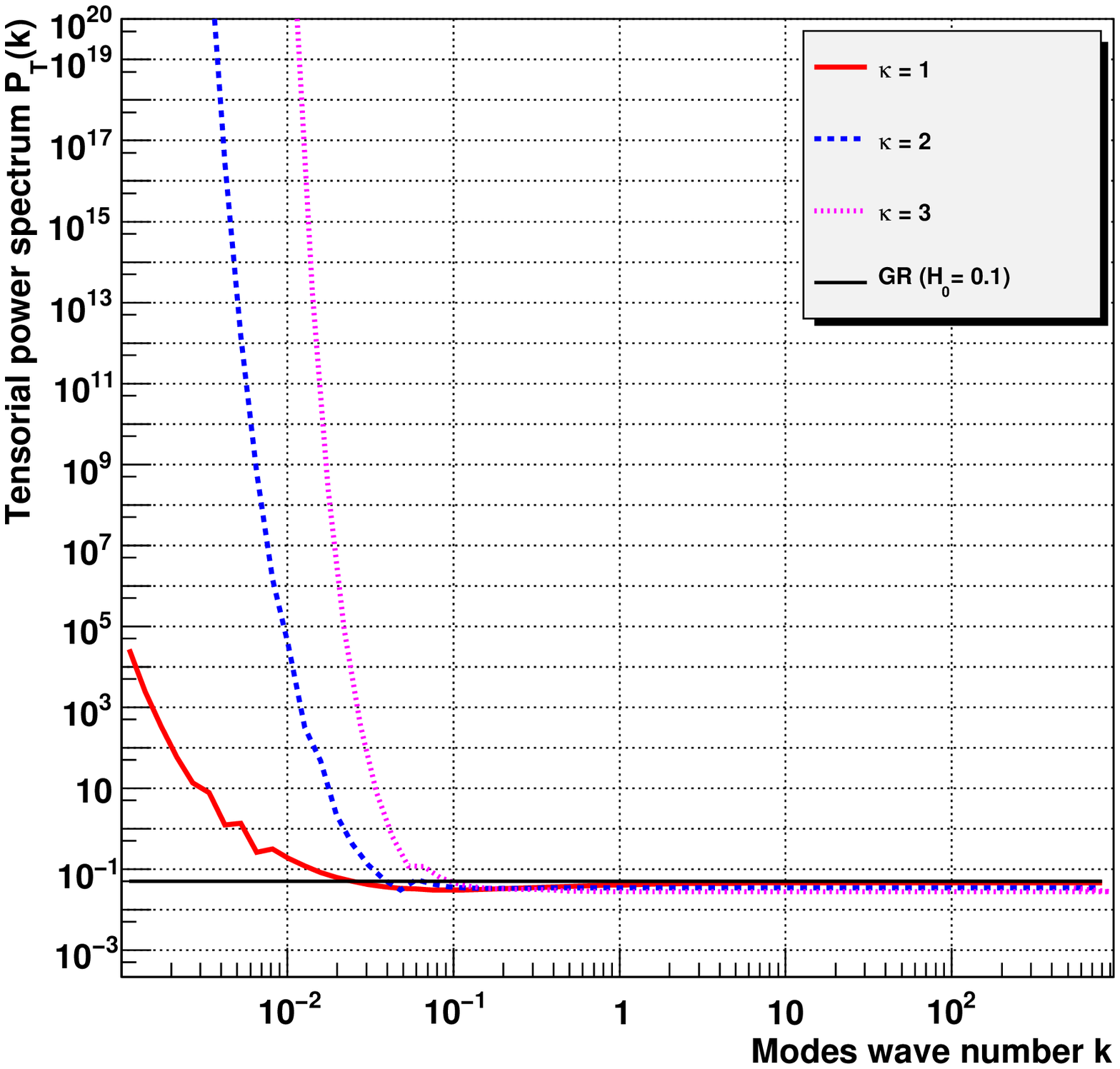}
	\caption{Primordial tensor power spectrum as a function of the wavenumber for different
	values of the LQC parameters $\lambda$ (left panel) and $\kappa$ (right panel).}
	\label{fig2}
	\end{center}
\end{figure*}


\begin{figure}
	\begin{center}
	\includegraphics[scale=0.45]{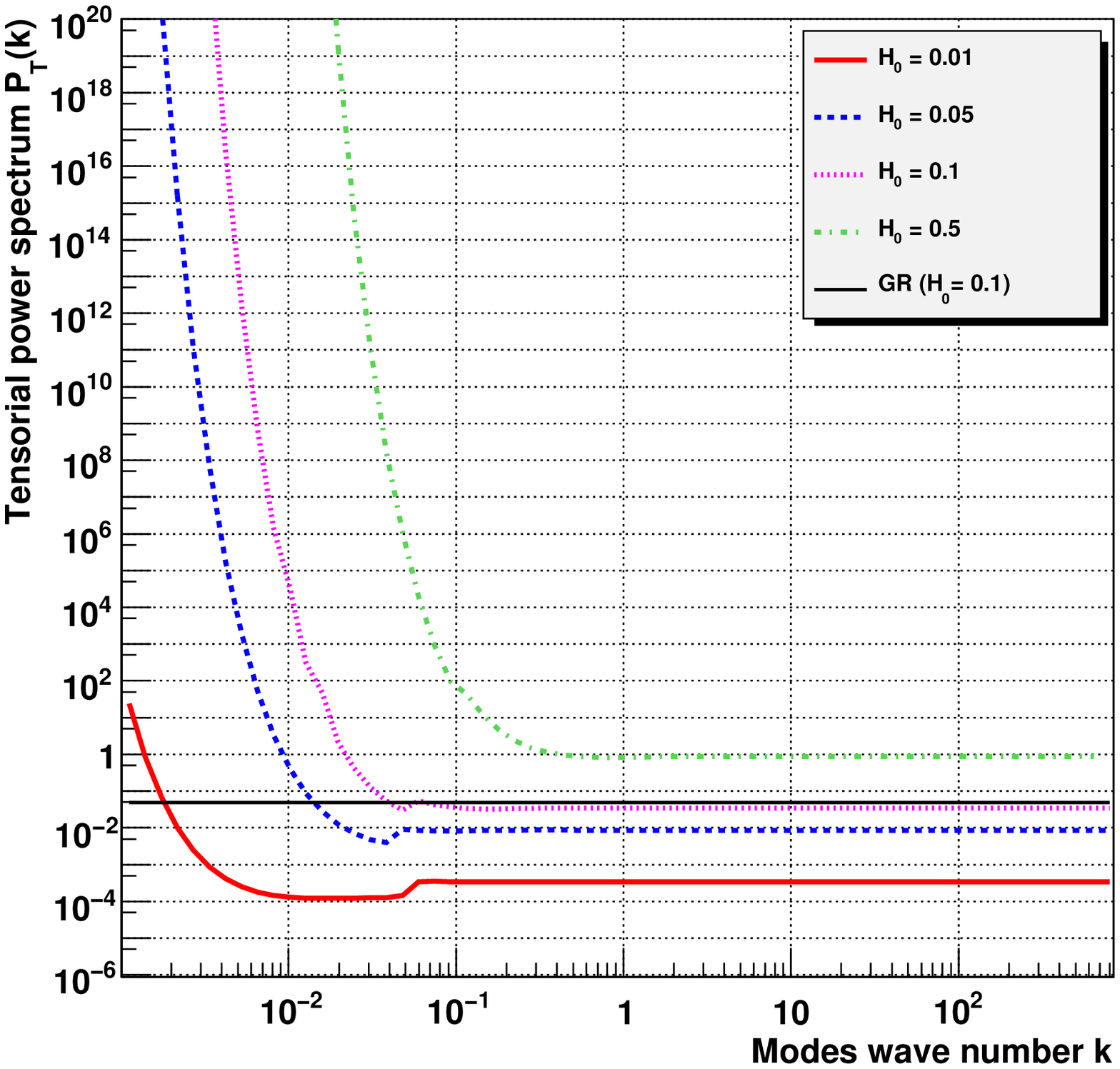}
	\caption{Primordial tensor power spectrum as a function of the wavenumber for different
	values of the parameter $H_0$.}
	\label{fig4}
	\end{center}
\end{figure}

A few comments can be made about the general
structure of Eq.~(\ref{equmain}), with the energy and potential terms given by Eq.~(\ref{elqg}) and 
Eq.~(\ref{vlqg}). First, it should be noticed that in the limit $\eta \to -\infty$, the potential is
always subdominant when compared with the energy. This makes meaningful the assumption of "asymptotically free"
initial states. However, in the remote past, the LQC correction terms are dominant, both in the
energy and in the potential. This is why the vacuum structure in the limit  $\eta \to -\infty$ is
far from obvious and the intuitive interpretation of the result is very difficult. Furthermore, 
if $\kappa (1+\epsilon)>2$ the potential diverges when $\eta \to -\infty$. This unusual feature
remains harmless as the energy term is, whatever the parameters, much higher than the
potential term. When  $\eta \to 0$, {\it i.e.} at the end of inflation, both the energy and the
potential are dominated by the GR terms.



Figures~\ref{fig1}, Fig.~\ref{fig2} and Fig.~\ref{fig4} display the primordial tensor 
power spectrum as a function of the wavenumber for different values of the physical parameters.
Our fiducial model is defined by $\epsilon=0$, $\lambda=1$, $\kappa=2$, $H_0=0.1$. We remind that
$\epsilon$ controls the slow-roll whereas $\lambda$ and $\kappa$ are LQC parameters defined by 
Eq.~(\ref{d}) and $H_0$ is such that $H_0\equiv\ell_{Pl}/\ell_0$ with 
$a(\eta)=\ell_0\left|\eta\right|^{-1-\epsilon}$.
The main result of this article, which appears in all the figures, can easily be
noticed: the spectrum in strongly infrared divergent due to LQC corrections. On the other hand, as
expected, the UV limit coincides with the GR prediction.

Fig.~\ref{fig1} aims at underlining the general evolution tendency of the spectrum as a function of
the slow-roll parameter and the last value ($\epsilon$=0.1) is deliberately chosen above the
observationally allowed range \cite{wmap}.
Although it cannot be easily seen on the plot due to the scale, the spectrum exhibits,
as it should, a $-2\epsilon$ tilt in the UV limit. On Fig.~\ref{fig2}, one can
notice that, as expected, the higher the value of the LQC parameters, the stronger the deviation
from GR. Finally, the case of Fig.~\ref{fig4} is slightly more intricate as $H_0$ is fundamentally
a background parameters which, however, couples to LQC corrections {\it via} the $\lambda$ term in
the IR regime.



\begin{figure*}
	\begin{center}
	\includegraphics[scale=0.4]{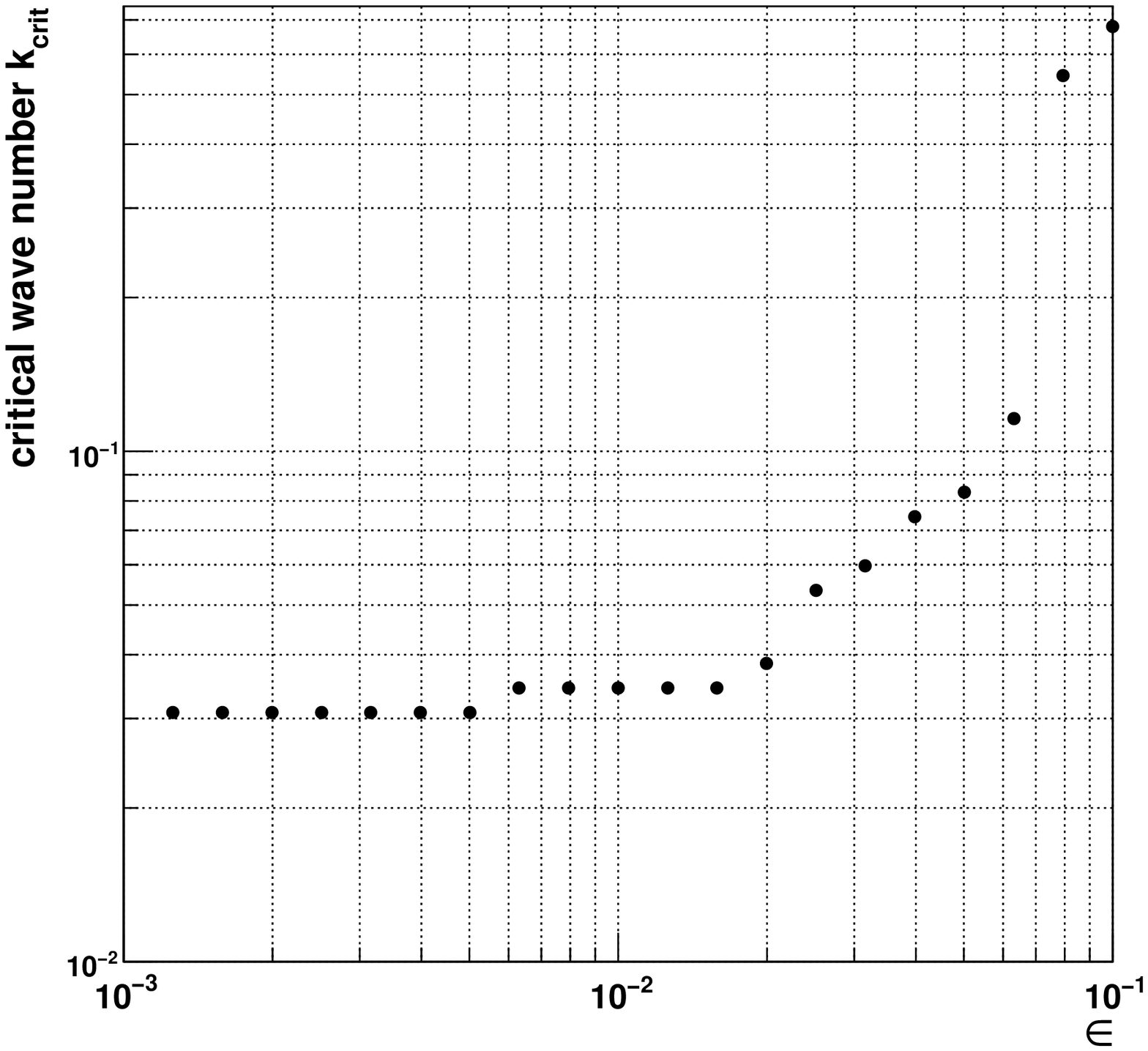}
	\includegraphics[scale=0.4]{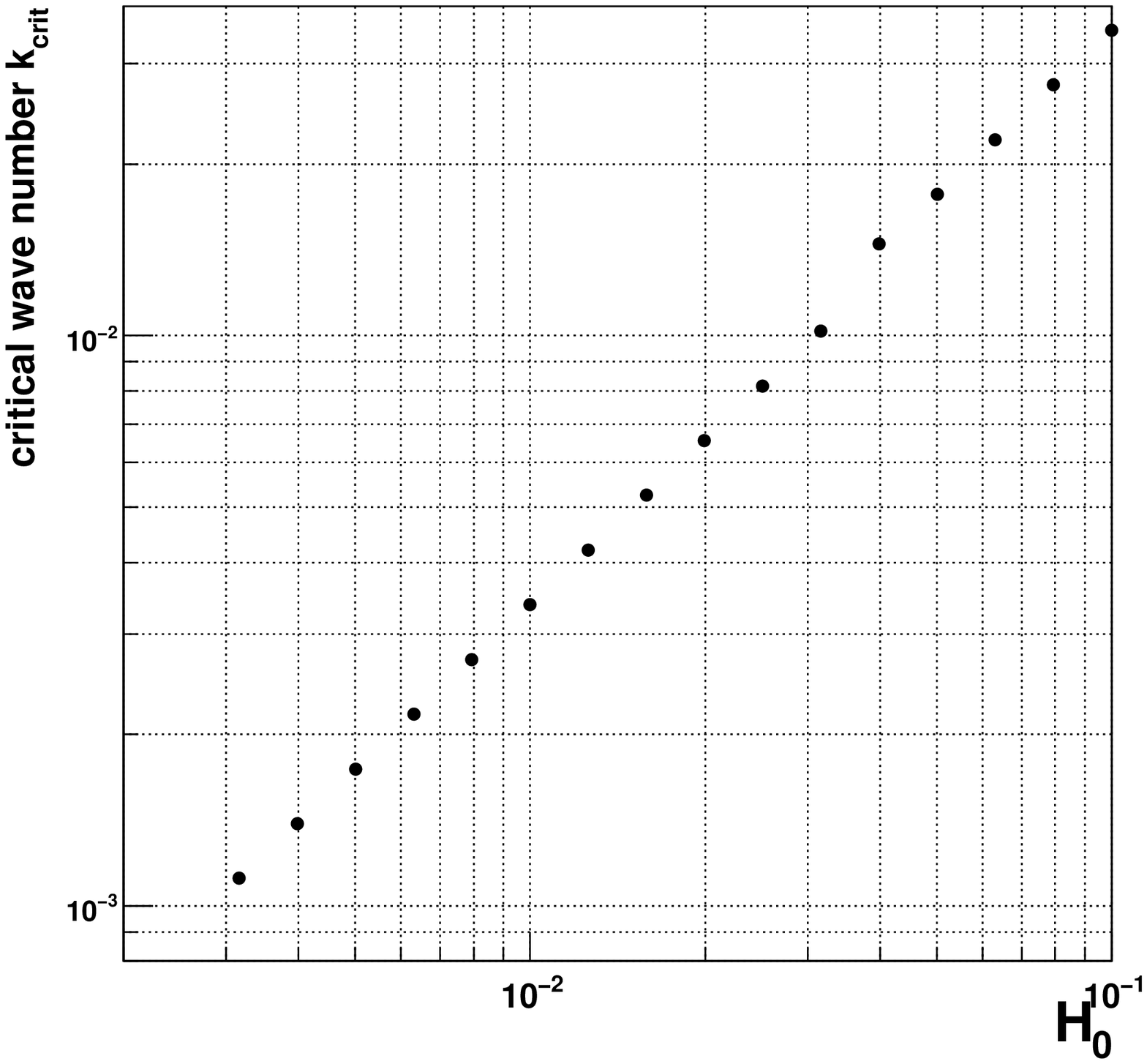} \\
	\includegraphics[scale=0.4]{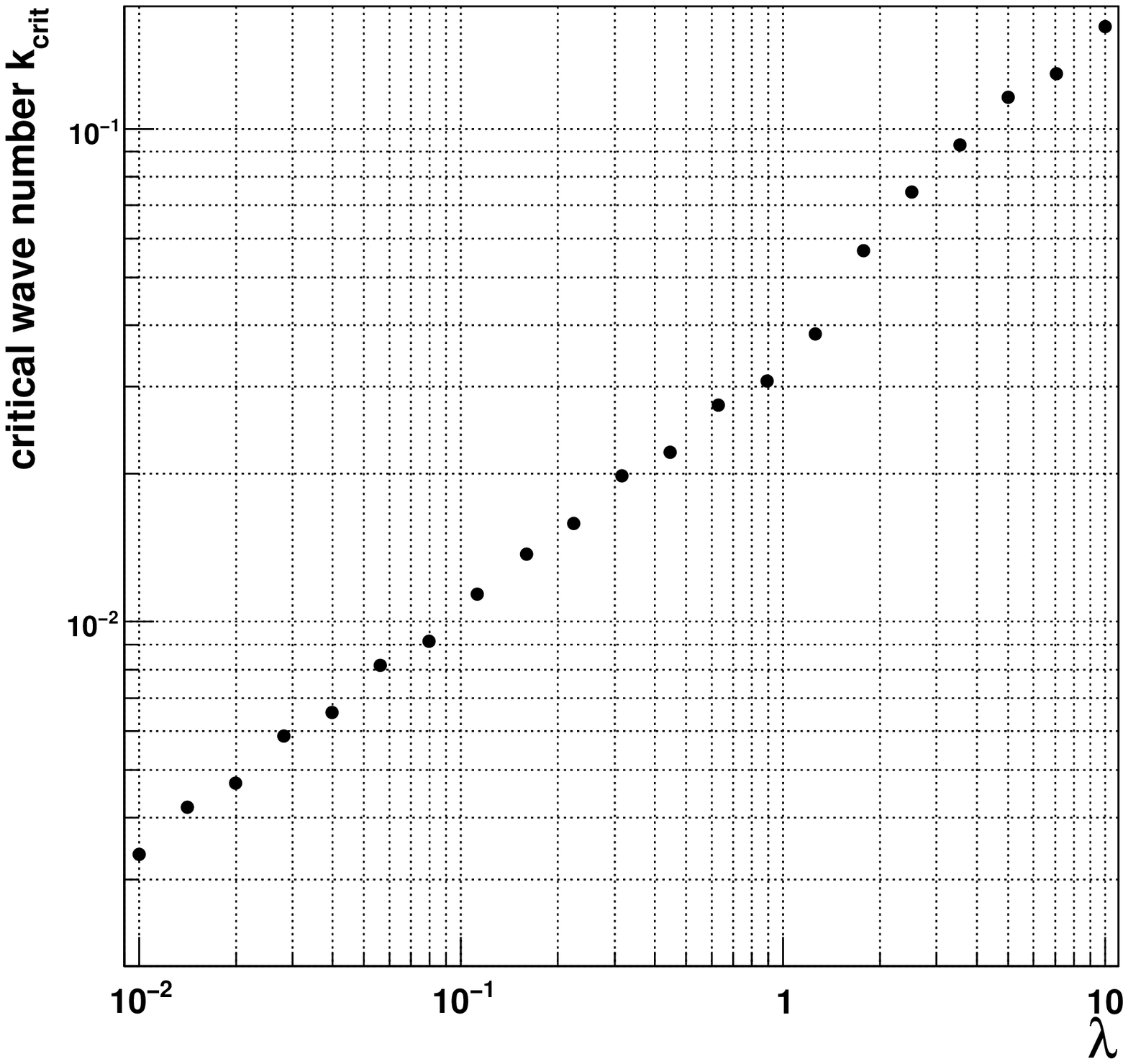}
	\includegraphics[scale=0.4]{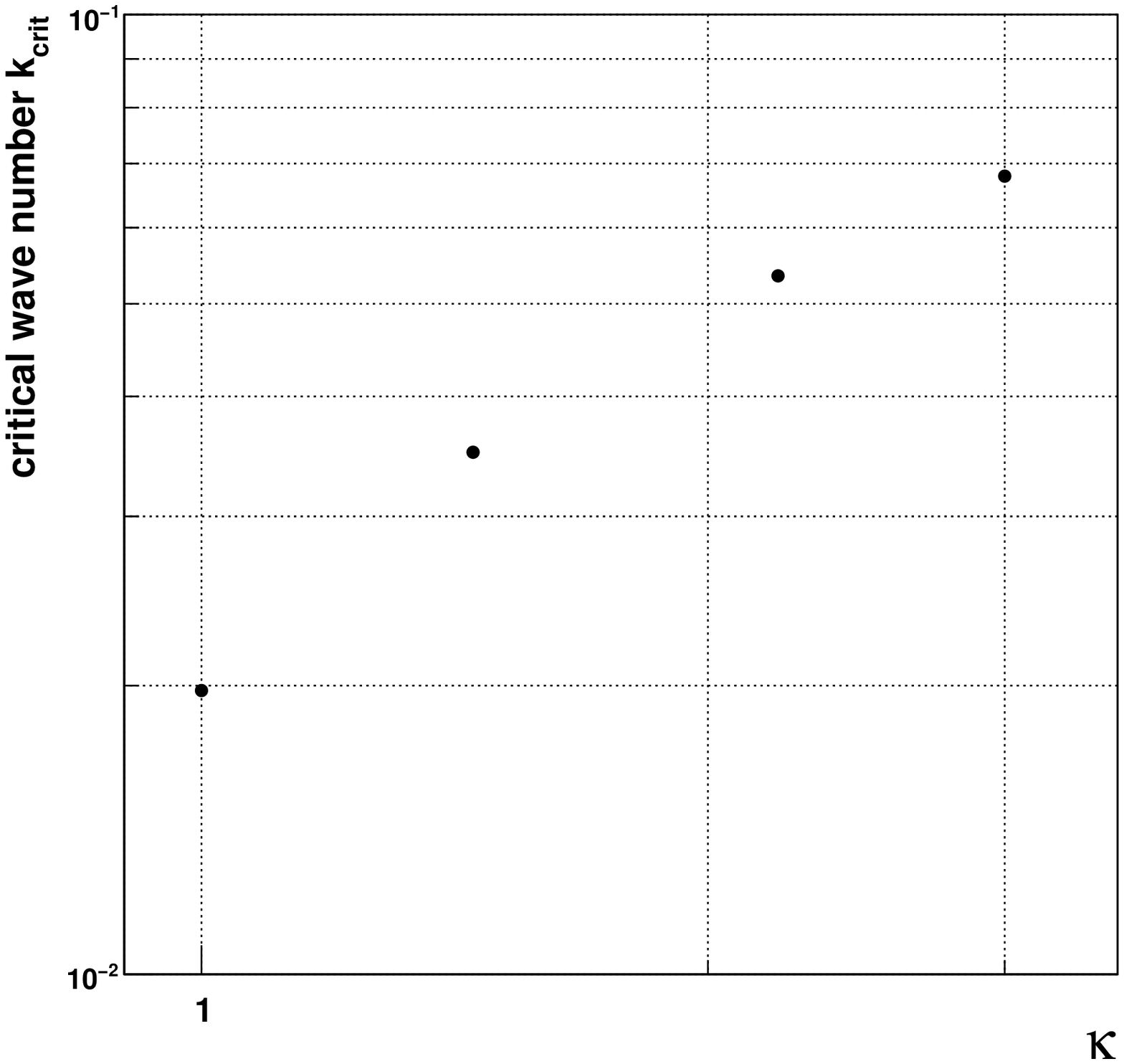}
	\caption{Value of the critical wavenumber as a function of the inflationary parameters $\epsilon$ (top-left) and $H_0$ (top-right) and as a function of the LQC parameters $\lambda$ (bottom-left) and $\kappa$ (bottom-right).}
	\label{fig5}
	\end{center}
\end{figure*}

On Fig.~\ref{fig5}, the critical wavenumber
$k_{crit}$ is displayed as a function of the parameters. We define $k_{crit}$ so that
$\mathcal{P}_{\mathrm{T}}(k_{crit})=2lim_{k\to \infty}\mathcal{P}_{\mathrm{T}}(k)$ (except for a nonvanishing $\epsilon$, in which case the criterion is $\mathcal{P}^{(LQG)}_{\mathrm{T}}(k_{crit})=2\mathcal{P}^{(RG)}_{\mathrm{T}}(k_{crit})$), as an indicator of the transition wavenumber 
between the LQC-dominated ($k<k_{crit}$)
and the GR-dominated ($k>k_{crit}$) regimes. Interestingly, it can be concluded from those plots
that this critical wavenumber is highly dependent upon any LQC parameter. To some
extent, this feature is more observationally relevant than the amplitude of the
effect which is anyway huge in the infrared limit. The higher the LQC correction, the
higher the critical wavenumber, the smaller the physical scales submitted to LQC
corrections, and the easier the observation. Although probably fortuitous, it is worth noticing that
the transition scale is nearly proportional to the energy scale of inflation. 
The dependence upon the first slow-roll parameter is very weak in the allowed
range, making the predictions reliable from the viewpoint of a test of LQC.

\section*{CONCLUSION}
	
The influence of holonomy corrections during slow-roll inflation was derived in 
\cite{grainlqg1,grainlqg2}. This article follows the same approach but considers inverse-volume
terms (complementing the approach of \cite{calcagni} which was performed in the framework of superinflation). Both analytical results (for a few
particular cases) and  numerical results (sampling the full parameter space) were obtained. The general
behavior is a very substantial deviation from GR in the $k\to 0$ limit. This deviation
affects the amplitude of the power spectrum and, more importantly, both the 
tilt and the running of the index. There are several ways
in which this work should be developed. First, it would be welcome to include simultaneously 
holonomy and inverse-volume corrections in the same differential equation for primordial 
gravitons. Then,
and more importantly, corrections to the background should also be taken into account. Although
several articles are devoted to this point, no fully consistent numerical study of LQC
propagation {\it and} background corrections is yet available. Finally,
it would be interesting to investigate LQG corrections to the scalar spectrum \cite{bojoscal}.
This last point is probably the most promising one from the observational viewpoint.

\end{document}